\documentclass[useAMS,usenatbib,usegraphicx]{mn2e}
\usepackage{times}
\usepackage{epsfig}
\title[The most massive star cluster in the Local Group]{A ``super'' star
cluster grown old: the most massive star cluster in the Local Group}

\author[J. Ma et al.]
{J. Ma,$^1$\thanks{E-mail: majun@vega.bac.pku.edu.cn}
R. de Grijs,$^2$
Y. Yang,$^1$
X. Zhou,$^1$ J. Chen,$^1$ Z. Jiang,$^1$ Z. Wu$^1$ and J.Wu$^1$\\
$^1$National Astronomical Observatories, Chinese Academy of
Sciences, 20A Datun Road, Chaoyang District, Beijing 100012, China\\
$^2$Department of Physics \& Astronomy, The University of Sheffield,
Hicks Building, Hounsfield Road, Sheffield S3 7RH
}

\date{Received 2006 Jan 16; Accepted 2006 Feb 22}

\pubyear{2006}

\begin{document}

\label{firstpage}

\maketitle

\begin{abstract}
We independently redetermine the reddening and age of the globular
cluster 037-B327 in M31 by comparing independently obtained
multicolour photometry with theoretical stellar population synthesis
models. 037-B327 has long been known to have a very large reddening
value, which we confirm to be $E(B-V)=1.360\pm0.013$, in good
agreement with the previous results. We redetermine its most likely
age at $12.4\pm 3.2$ Gyr.\\
037-B327 is a prime example of an unusually bright early
counterpart to the ubiquitous ``super'' star clusters presently
observed in most high-intensity star-forming regions in the local
Universe. In order to have survived for a Hubble time, we conclude
that its stellar IMF cannot have been top-heavy. Using this
constraint, and a variety of SSP models, we determine a
photometric mass of $\mathcal{M}_{\rm GC} = (3.0 \pm 0.5) \times
10^7$ M$_\odot$, somewhat depending on the SSP models used, the
metallicity and age adopted and the IMF representation.  This
mass, and its relatively small uncertainties, make this object the
most massive star cluster of any age in the Local Group. Assuming
that the photometric mass estimate thus derived is fairly close to
its dynamical mass, we predict that this GC has a
(one-dimensional) velocity dispersion of order $(72 \pm 13)$ km
s$^{-1}$. As a surviving ``super'' star cluster, this object is of
prime importance for theories aimed at describing massive star
cluster evolution.
\end{abstract}


\section{Introduction}

Among the Local Group of galaxies, M31 contains the largest
population of globular clusters (GCs) and is the nearest analogue
of the Milky Way. From the observational evidence collected thus
far \citep[see, e.g.,][]{rich05}, the M31 GCs reveal some striking
similarities to their Galactic counterparts
\citep{ffp94,dj97,bhh02}. For example, both GC systems seem to
have similar mass-to-light ratios, velocity dispersion --
luminosity relations \citep[see also][]{degrijs05} and structural
parameters.

Since the pioneering work of \citet{Tinsley68,Tinsley72} and
\citet{SSB73}, evolutionary population synthesis modeling has become a
powerful tool to interpret integrated spectrophotometric observations
of galaxies and their subcomponents, such as star clusters
\citep{Anders04}. Such models, as e.g. developed by \citet{bc93,bc96},
\citet{Leitherer95}, and \citet{FR97}, were comprehensively compiled
by \citet{Leitherer96} and \citet{Kennicutt98}. The evolution of star
clusters is usually modeled by means of the simple stellar population
(SSP) approximation. An SSP is defined as a single generation of
coeval stars formed from the same progenitor molecular cloud (thus
implying a single metallicity), and governed by a given initial mass
function (IMF). GCs, which are bright and easily identifiable, and
whose populations in a given galaxy are typically characterized by
homogeneous abundance and age distributions, are relatively easy to
understand compared to the other mix of stellar populations in
galaxies. \citet{bh01} compared the predicted SSP colours of three
stellar population synthesis models to the intrinsic broad-band
$UBVIRJHK$ colours of Galactic and M31 GCs, and found that the
best-fitting models match the clusters' spectral energy distributions
(SEDs) very well indeed. So, from the results of \citet{bh01}, there
is evidence that the stellar population of GCs may be described by the
SSPs of stellar population synthesis models. In fact, many authors
have used SSP models to study the populations of clusters across the
entire age range. For example, \citet{degrijs03b} simultaneously
obtained ages, metallicities and extinction values for ~300 clusters
in NGC~3310, based on archival {\sl Hubble Space Telescope (HST)}
observations from the ultraviolet to the near-infrared by means of a
comparison between the observed SEDs and the predictions from the {\sc
galev} SSP models \citep{schulz02,afva03} \citep[see
also][]{degrijs03c}. Using their sophisticated and extensively
validated method, \citet{degrijs03a} obtained the age and mass
estimates for 113 star clusters in the fossil starburst region B of
M82 by comparing the observed cluster SEDs with the model predictions
for an instantaneous burst of star formation. \citet{bik03} and
\citet{bastian05} derived ages, initial masses and extinctions of M51
star cluster candidates by fitting {\sc Starburst99} SSP models
\citep{Leitherer99} for instantaneous star formation to the observed
SEDs based on {\sl HST}/WFPC2 observations in six broad-band and two
narrow-band filters.  \citet{Ma01,Ma02a,Ma02b,Ma02c} and
\citet{jiang03} estimated the ages of 180 star clusters in M33 and 172
GC candidates in M31 by comparing the SSP synthesis models of BC96
\citep{bc96} with the integrated photometric measurements of these
objects in the Beijing-Arizona-Taiwan-Connecticut (BATC) photometric
system.

The study of M31 has been, and continues to be, a corner stone in
extragalactic astrophysics \citep{bh00}. The study of GCs in M31
can be traced back to \citet{hubble32}, who discovered 140 GCs
with $m_{\rm pg}\leq 18$ mag. In this paper, we discuss the
properties of the M31 GC B327 (where B indicates `Baade') or Bo037
(Bo = `Bologna', see Battistini 1987), which will subsequently be
referred to as 037-B327, following the nomenclature introduced by
\citet{huchra91}. The extremely red colour of this object,
combined with its apparent magnitude (given in their paper as $V_0
= 16.97$ mag), was first noted by \citet{kronmay60}, who suggested
that this implied that the cluster must be highly reddened ($A_V =
2.35$ mag) and extremely luminous, $M_V^0 \simeq -11.5$ mag. Two
years later, \citet{vete62a} determined magnitudes of 257 GC
candidates in M31, including 037-B327, in the $UBV$ photometric
system. Using his photometric catalog, \citet{vete62b} studied the
intrinsic colours of the M31 GCs, and found that 037-B327 was the
most highly reddened objects in his sample of M31 GC candidates,
with $E(B-V)=1.28$ mag. In order to avoid any {\it a priori}
reason implying that the intrinsically brightest GCs in M31 should
also be the most highly reddened, \citet{Sidney68,Sidney69} argued
that $R_V \equiv A_V/E(B-V)<3.0$ in M31. Based on a re-analysis of
the reddening towards the M31 GC population in general, and to
037-B327 in particular, \citet{bk02} argued that the evidence for
an unusual reddening law is `somewhat less compelling' than
implied by \citet{Sidney68}'s arguments.  Using low-resolution
spectroscopy, \citet{cram85} also found this cluster to be the
most highly reddened GC candidate in M31, with $E(B-V)=1.48$ mag.
Armed with a large database of multicolour photometry,
\citet{bh00} determined the reddening value for each individual
M31 GC, including 037-B327, using the correlations between optical
and infrared colours and metallicity based on various
``reddening-free'' parameters, and found $E(B-V)=1.38\pm0.02$ mag
for 037-B327 (P. Barmby, priv. comm.). Using the spectroscopic
metallicity to predict the intrinsic colours, \citet{bk02}
rederived the reddening value for this GC, $E(B-V)=1.30\pm0.04$
mag. Although the reddening values of 037-B327 based on a variety
of methods are consistent, this value is unusually large and
therefore worth verifying using independent methods. At the same
time, the large reddening value makes 037-B327 the most {\it
intrinsically} luminous GC in M31 \citep[see details
from][]{bk02}.

In this paper, we first redetermine the reddening for 037-B327 by
comparing observational SEDs (Sect. \ref{data.sec}) with population
synthesis models in Sect. \ref{parameters.sec}. Our independently
determined results are in very good agreement with previous
determinations. In Sect. \ref{interpretation.sec} we then place this
cluster in its evolutionary context, and conclude that 037-B327 is, in
fact, not only a surviving ``super'' star cluster, but also the most
massive cluster of any age known in the Local Group. We summarize our
results in Sect. \ref{summary.sec}.

\section{BATC, Broad-band and 2MASS Photometry of 037-B327}
\label{data.sec}

\subsection{Archival Images of the BATC Sky Survey}

\begin{figure}
\includegraphics[width=8.4cm]{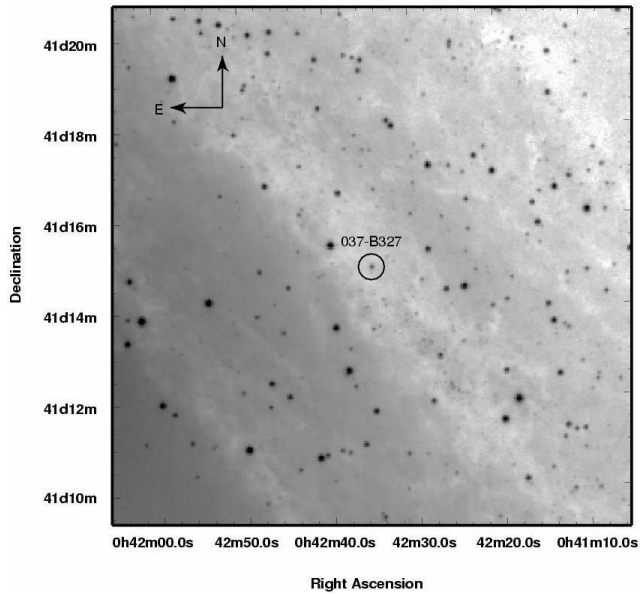}
\caption{An image of 037-B327 in the BATC $g$ band obtained with the
NAOC 60/90cm Schmidt telescope. 037-B327 is circled. The field of view
of the image is $11^{\prime}\times 11^{\prime}$.}
\label{chart.fig}
\end{figure}

The observations of M31 were obtained by the
Beijing-Arizona-Taiwan-Connecticut (BATC) 60/90cm f/3 Schmidt
telescope located at the XingLong station of the National
Astronomical Observatory of China (NAOC). This telescope is
equipped with a Ford Aerospace $2048\times 2048$ CCD camera with
15 $\mu$m pixel size, giving a CCD field of view of $58^{\prime}$
$\times $ $58^{\prime}$ with a pixel size of
$1\arcsec{\mbox{}\hspace{-0.15cm}.} 7$. The BATC survey was
carried out using 15 intermediate-band filters covering the
optical wavelength range, $\sim 3000-10000$\AA. These filters were
specifically designed to avoid contamination from the brightest
and most variable night-sky emission lines. A description of the
BATC photometric system can be found in \citet{fan96}. The finding
chart of 037-B327 in the BATC $g$ band (centered on 5795\AA),
obtained with the NAOC 60/90cm Schmidt telescope, is shown in Fig.
\ref{chart.fig}.

\subsection{Intermediate-Band Photometry of 037-B327}

\citet{jiang03} extracted 123 images of M31 from the BATC survey
archive, taken in 13 BATC filters (excluding the $a$ and $b$ filters)
between 1995 September and 1999 December, and combined multiple images
of the same filter to improve the image quality. Subsequently, they
determined the magnitudes of 172 GCs, including 037-B327, in these 13
BATC filters based on the combined images using standard aperture
photometry, i.e. essentially by employing the PHOT routine in {\sc
daophot} \citep{stet87}. The BATC photometric system calibrates the
magnitude zero level in a similar fashion to the spectrophotometric AB
magnitude system. For the flux calibration, the Oke-Gunn primary flux
standard stars HD 19445, HD 84937, BD +26$^{\circ}$2606, and BD
+17$^{\circ}$4708 \citep{ok83}, were observed during
photometric nights \citep{yan00}. Table 1 lists the resulting BATC
photometry of 037-B327.

\begin{table}
\caption{BATC Photometry of the M31 GC 037-B327.}
\begin{center}
\begin{tabular}{ccccc}
\hline Filter & $\lambda_{\rm central}$ (\AA) & FWHM (\AA) & $N^a$& Magnitude \\
\hline
 c & 4210 & 320 &  3 & 19.28(0.061)\\
 d & 4540 & 340 & 17 & 18.47(0.032)\\
 e & 4925 & 390 & 11 & 17.81(0.021)\\
 f & 5270 & 340 & 12 & 17.23(0.016)\\
 g & 5795 & 310 &  7 & 16.40(0.012)\\
 h & 6075 & 310 &  5 & 16.14(0.009)\\
 i & 6656 & 480 &  3 & 15.54(0.007)\\
 j & 7057 & 300 & 12 & 15.25(0.006)\\
 k & 7546 & 330 &  6 & 14.89(0.006)\\
 m & 8023 & 260 & 12 & 14.61(0.004)\\
 n & 8480 & 180 &  5 & 14.32(0.007)\\
 o & 9182 & 260 & 18 & 13.95(0.004)\\
 p & 9739 & 270 & 12 & 13.78(0.005)\\
\hline
\multicolumn{5}{l}{$^a$ $N$ is the number of images taken
by the BATC survey.}\\
\end{tabular}
\end{center}
\end{table}

\subsection{Broad-band and 2MASS Photometry of 037-B327}

In order to estimate the reddening value of 037-B327 accurately, we
try to use as many photometric data points covering as large a
wavelength range as possible. Using the 4-Shooter CCD mosaic camera
and the SAO infrared imager on the 1.2m telescope at the Fred Lawrence
Whipple Observatory (FLWO), \citet{bh00} presented optical and
infrared photometric data for 285 M31 GCs \citep[see Table 3
of][]{bh00}. However, for 037-B327, only photometric measurements
through the $BVRI$ filters were listed. \citet{sl81} presented the
photoelectric $UBV$ photometry of 110 M31 GCs, including 037-B327,
with the 6m telescope of the Special Astrophysical Observatory. We
adopted their photometric data point in the $U$ band, with a
photometric uncertainty of 0.08 mag, as suggested by \citet{gall04}.

Using the 2MASS database, \citet{gall04} identified 693 known and
candidate GCs in M31, and presented their 2MASS $JHK$s magnitudes
[\citet{gall04} transformed all 2MASS magnitudes to the CIT
photometric system \citep{Elias82,Elias83} using the colour
transformations in \citet{Carpenter01}]. In this paper, we need the
2MASS $JHK$s magnitudes for 037-B327 in order to compare our
observational SED to the SSP models, so we reversed this
transformation using the same procedures. Since \citet{gall04} did not
give the 2MASS $JHK$s photometric uncertainties, we adopted 0.03 mag
for all of the $J$, $H$, and $K$s bands. We obtained these uncertainty
estimates by comparing the photometry of 037-B327 with Fig. 2 of
\citet{Carpenteretal01} who plotted the observed photometric rms
uncertainties in the time series as a function of magnitude for stars
brighter than their observational completeness limits. In fact, the
photometric uncertainties adopted do not affect our results
significantly. The broad-band and 2MASS photometric data points of
037-B327 are listed in Table 2.

\begin{table}
\caption{Broad-band and 2MASS Photometry of the M31 GC 037-B327.}
\begin{center}
\begin{tabular}{lll}
\hline
  Filter & Magnitude   & Reference\\
\hline
   $U$   & 20.10(0.08) & \citet{sl81}\\
         &             &             \\
   $B$   & 18.87(0.05) & \citet{bh00}\\
   $V$   & 16.82(0.05) &             \\
   $R$   & 15.54(0.05) &             \\
   $I$   & 14.19(0.05) &             \\
         &             &             \\
   $J$   & 12.24(0.03) & \citet{gall04}\\
   $H$   & 11.26(0.03) &               \\
   $K_{\rm s}$ & 11.02(0.03) &               \\
\hline
\end{tabular}
\end{center}
\end{table}

\section{Parameter determinations for 037-B327}
\label{parameters.sec}

\subsection{Stellar Populations and Synthetic Photometry}

To estimate the reddening and age of 037-B327, we compare its SED with
theoretical stellar population synthesis models. We used the SSP
models of \citet[hereafter BC03]{bc03}, which have been upgraded from
the \citet{bc93,bc96} version, and now provide the evolution of the
spectra and photometric properties for a wider range of stellar
metallicities. BC03 provide 26 SSP models (both of high and low
resolution) using the 1994 Padova evolutionary tracks, 13 of which
were computed using the \citet{chabrier03} IMF assuming lower and
upper mass cut-offs of $m_{\rm L}=0.1$ M$_{\odot}$ and $m_{\rm U}=100$
M$_{\odot}$, respectively, while the other 13 were computed using the
\citet{salpeter55} IMF with the same mass cut-offs. In addition, BC03
provide 26 SSP models using the 2000 Padova evolutionary
tracks. However, as \citet{bc03} pointed out, the 2000 Padova models,
which include more recent input physics than the 1994 models, tend to
produce worse agreement with observed galaxy colours. These SSP models
contain 221 spectra describing the spectral evolution of an SSP from 0
to 20 Gyr. The evolving spectra include the contribution of the
stellar component in the range from 91\AA~~to $160\mu$m. In this
paper, we adopt the high-resolution SSP models computed using the 1994
Padova evolutionary tracks and a \citet{salpeter55} IMF\footnote{We
note that because of the slow SED evolution of SSPs at ages in excess
of a few Gyr, all of the most commonly used spectral synthesis models
agree very well at these ages. Therefore, the choice of IMF is {\it
only} important for estimating the photometric mass of the cluster
(which we will discuss in Sect. \ref{interpretation.sec}), and does
{\it not} affect the determination of the age and reddening parameters
of 037-B327.}.

We convolve the SSP SEDs from BC03 with the BATC
intermediate-band, broad-band $UBVRI$ and 2MASS filter response
functions to obtain synthetic optical and near-infrared photometry
for comparison. The synthetic $i^{\rm th}$ filter magnitude can be
computed by
\begin{equation}
m=-2.5\log\frac{\int_{\lambda}F_{\lambda}\varphi_{i} (\lambda){\rm
d}\lambda}{\int_{\lambda}\varphi_{i}(\lambda){\rm d}\lambda}-48.60
\quad,
\end{equation}
where $F_{\lambda}$ is the theoretical SED and $\varphi_{i}$ the
response function of the $i^{\rm th}$ filter of the BATC, $UBVRI$ and
2MASS photometric systems. Here, $F_{\lambda}$ varies with age and
metallicity.

\subsection{Metallicity of 037-B327}

The SEDs of clusters are significantly affected by the metallicity
one adopts. Using the Wide Field Fibre Optic Spectrograph at the
William Herschel 4.2-m telescope, \citet{perr02} obtained spectra
of over 200 M31 GCs, including 037-B327. They determined a
metallicity for 037-B327 of ${\rm [Fe/H]}=-1.07\pm 0.20$ (or $Z =
0.0017$), which we adopt here.

\subsection{Fit Results}

We use a $\chi^2$ minimization test to examine which BC03 SSP
models are most compatible with the observed SED, following
\begin{equation}
\chi^2=\sum_{i=1}^{21}{\frac{[m_{\lambda_i}^{\rm
intr}(E(B-V))-m_{\lambda_i}^{\rm mod}(t)]^2}{\sigma_{i}^{2}}}
\quad,
\end{equation}
where $m_{\lambda_i}^{\rm mod}(t)$ is the integrated magnitude in
the $i^{\rm th}$ filter of an SSP at age $t$, $m_{\lambda_i}^{\rm
intr}(E(B-V))$ presents the intrinsic integrated magnitude in the
same filter, and
\begin{equation}
\sigma_i^{2}=\sigma_{{\rm obs},i}^{2}+\sigma_{{\rm mod},i}^{2} \quad .
\end{equation}
Here, $\sigma_{{\rm obs},i}^{2}$ is the observational uncertainty, and
$\sigma_{{\rm mod},i}^{2}$ is the uncertainty associated with the
model itself, for the $i^{\rm th}$ filter. \citet{charlot96} estimated
the uncertainty associated with the term $\sigma_{{\rm mod},i}^{2}$ by
comparing the colours obtained from different stellar evolutionary
tracks and spectral libraries. Following \citet{wu05}, we adopt
$\sigma_{{\rm mod},i}^{2}=0.05$ in this paper.

\begin{figure}
\includegraphics[width=8.4cm]{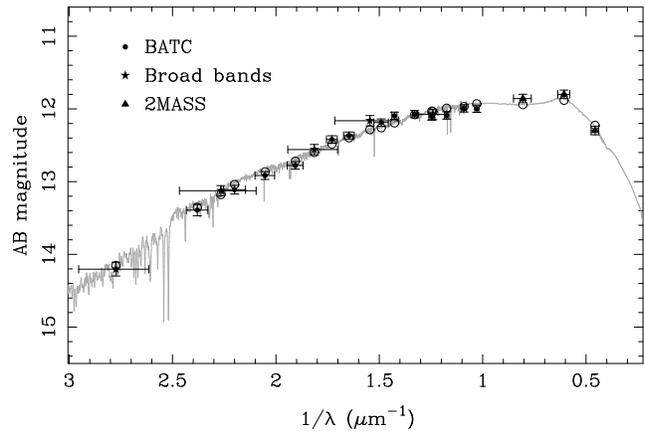}
\caption{Comparison of the best-fit integrated theoretical SED with
the intrinsic SED of 037-B327. The photometric data points are shown
as the symbols with error bars (vertical error bars for photometric
uncertainties and horizontal ones for the approximate wavelength
coverage of each filter). Open circles represent the calculated
magnitude of the model SED for each filter.}
\label{bestfit.fig}
\end{figure}

\begin{figure*}
\vspace{-3cm}
\includegraphics[width=\textwidth]{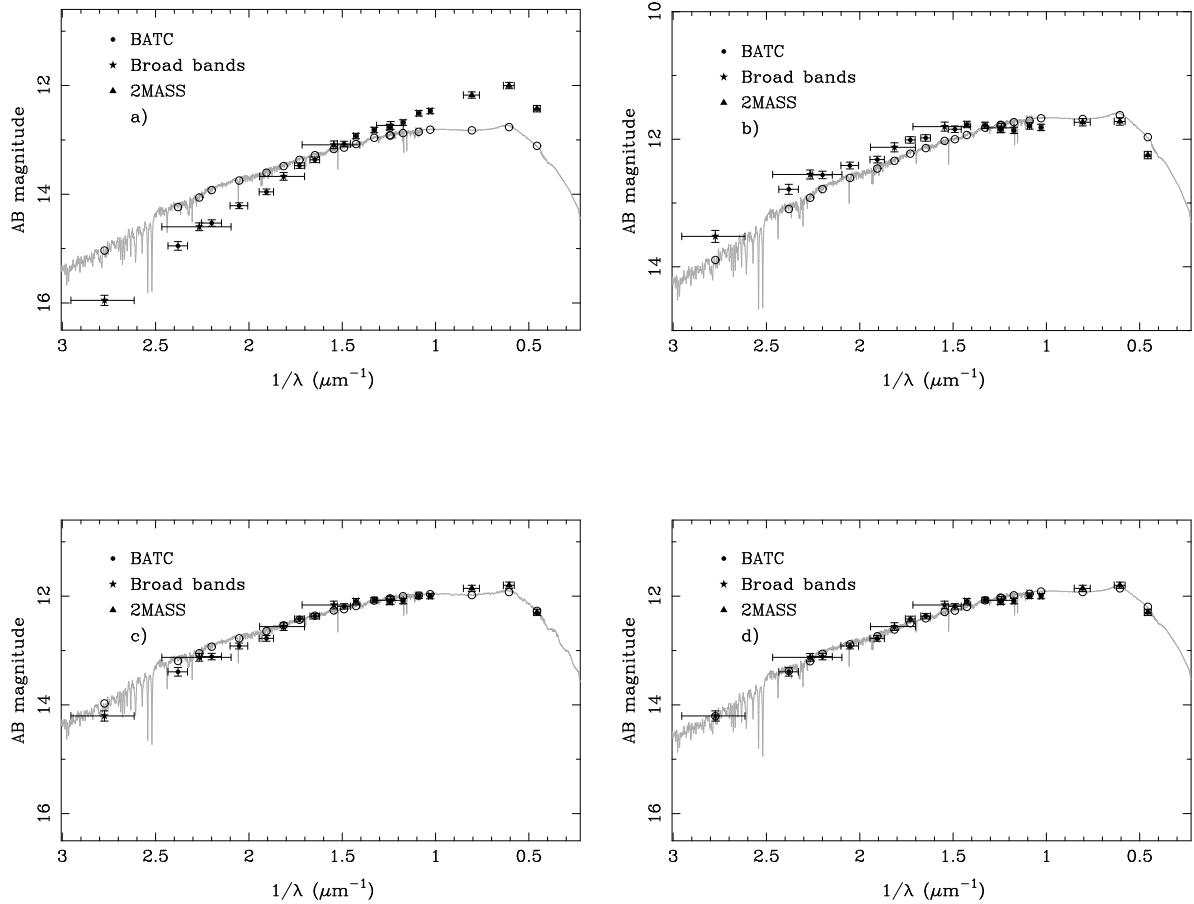}
\vspace{-6cm}
\caption{Comparison of the observed SED of 037-B327 with theoretical
SEDs covering a range in ages and reddening values. The symbols are as
in Fig. \ref{bestfit.fig}. (a) for an age of 12.4 Gyr, and $E(B-V) =
1.00$ mag; (b) as (a), but for $E(B-V)=1.50$ mag; (c) for
$E(B-V)=1.36$ mag, and an age of 5.0 Gyr; (d) as (c), but for an age of
17.0 Gyr.}
\label{fitranges.fig}
\end{figure*}

\begin{figure}
\resizebox{\hsize}{!}{\rotatebox{-90}{\includegraphics{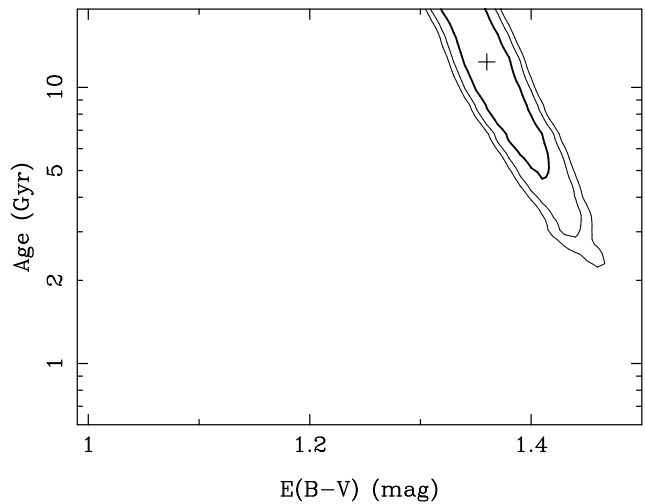}}}
\vspace{0.0cm} \caption{Likelihood contour plot in the age-reddening
plane for 037-B327, using BC03 models. The favoured solution has a
cluster age of 12.4 Gyr and a reddening of $E(B-V)=1.36$ mag. The
contours shown are for the 99.7, 95.4 and 68.3 per cent confidence
levels.}
\label{contours.fig}
\end{figure}

The BC03 SSP models include six initial metallicities, 0.0001,
0.0004, 0.004, 0.008, 0.02 (solar metallicity), and 0.05. Spectra
for other metallicities can be obtained by linear interpolation of
the appropriate spectra for any of these metallicities. We treat
$E(B-V)$ as a fit parameter, to be determined simultaneously with
the cluster age. The values for the extinction coefficient,
$R_{\lambda}$, are obtained by interpolating the interstellar
extinction curve of \citet{car89}. In Fig. \ref{bestfit.fig} we
show the intrinsic SED of 037-B327 and the SED of the best-fit
model. The best reduced $\chi^2_{\rm min}$ is achieved with a
reddening value of $E(B-V)=1.360\pm0.013$ mag and an age of
$12.4\pm{3.2}$~Gyr ($1\sigma$ uncertainties). The former is in
good agreement with previous results, i.e., $E(B-V)=1.28$ obtained
by \citet{vete62b}, $E(B-V)=1.38\pm0.02$ of \citet{bh00}, and
$E(B-V)=1.30\pm0.04$ of \citet{bk02}. Fig. \ref{contours.fig}
shows the contours of the 99.7, 95.4 and 68.3 per cent confidence
levels in the age-reddening plane for 037-B327. In order to
support our claim that the best fit in Fig. \ref{bestfit.fig} is
reasonable, in Fig. \ref{fitranges.fig} we show comparisons
between the observed SED of 037-B327 and theoretical SEDs covering
a range in ages and reddening values. It is clear that the
theoretical SEDs with the $E(B-V) = 1.00$ and $1.50$ mag do not
fit the observed SED of 037-B327 satisfactorily. Similarly, the
theoretical SED of 5.0 Gyr does not fit the observed SED of
037-B327 satisfactorily either, once again supporting our result
that this is an old globular cluster, at least older than a few
Gyr. On the other hand, theoretical SEDs with age of up to 17.0
Gyr would fit the observed SED of 037-B327 reasonably well. In
fact, the theoretical SEDs are not sensitive to the variation of
age for ages greater than $\sim 10$ Gyr. Thus, here we show
robustly that this cluster is as old as the majority of the
Galactic GCs.  Theoretical SEDs are sensitive to variations in
reddening, in particular when one has access to a large wavelength
coverage. Therefore, we conclude that the reddening value obtained
in this paper is robust. Deep observations of high spatial
resolution, taken with the Advanced Camera for Surveys on board
the {\sl HST} in the F606W filter show that 037-B327 is partially
crossed by a dust lane, which might be responsible for the bulk of
the reddening \citep{ma06}. The non-uniform optical depth across
the object may artifically redden its colours, with as a
consequence that the age obtained in this paper may be somewhat
older than the cluster's true age. However, since we cannot
disentangle these effects from the ground-based photometric
measurements owing to the high spatial resolution required for
this \citep{ma06}, the impact of the partial dust lane coverage
cannot be quantified at present. However, we point out that the
dust lane crosses the object in its periphery, and does not
significantly affect the bright core that dominates our integrated
photometry. As such, we believe that the importance of this
non-uniform reddening is limited. Nevertheless, we will take this
effect into account in Section \ref{interpretation.sec}, where we
will assume a lower age limit of 5 Gyr in order to validate the
robustness of our main results.

At the same time, however, \citet{ma06} derived that 037-B327 has
a high mean ellipticity, $\epsilon\simeq0.23$, which could either
imply that this cluster may not be very old, or that it has
recently been affected by significant tidal forces such as those
owing to a close encounter with a giant molecular cloud in the
disk of the galaxy (cf. N. Bastian \& S.P. Goodwin, in prep.). The
ellipticity of a cluster by itself cannot be taken as evidence of
a cluster being either young or old, however, although older
clusters are in general believed to be more spherical than younger
ones. For instance, in the Magellanic Clouds one finds
non-spherical clusters of all ages \citep{vdb91,goodwin97}, while
\citet{steph06} show for the case study of WLM-1, the lone GC
associated with the low-mass dwarf irregular galaxy WLM, that it
is highly elliptical (yet non-rotationally flattened) despite its
old age. We should keep in mind that although the age of 037-B327
obtained in this paper is model-dependent, independent studies
based on the photometry alone suggest an age of $\ga 10$ Gyr
\citep[e.g.][]{jiang03}, the range of which is encompassed by our
uncertainty estimate. Spectroscopic follow-up observations are
required to determine the cluster's age more conclusively.
Colour-magnitude analysis of the high-resolution ACS images will
not be able to improve this situation because of the very crowded
cluster field at the distance of M31.

\citet{jiang03} estimated the age of 037-B327 at 9.75~Gyr, based
on only the BATC data and on the SSP models of Bruzual \& Charlot
(1996; hereafter BC96); the reddening value adopted by
\citet{jiang03} was $E(B-V)=1.38\pm0.02$ \citep{bh00}.
\citet{jiang03} only used the BC96 models for three metallicities,
i.e., 0.0004, 0.004 and 0.02, and did not linearly interpolate to
find the best-fitting model. They adopted $Z=0.004$, significantly
more metal-rich than the metallicity of 0.0017 adopted in this
paper which obtained by \citet{perr02} from the spectra (see
details from section 3.2). For old GCs, the age/metallicity
degeneracy becomes important. As a consequence, we conclude that
the age of 037-B327 obtained by \citet{jiang03} is younger than
its true age. From the results presented in this paper, we
conclude that 037-B327 is as old as the majority of the Galactic
GCs.

\section{A ``super'' star cluster come of age?}
\label{interpretation.sec}

With the basic parameters of 037-B327 now firmly established after our
redetermination based on independent observational data, and on an
independent modelling approach, we can now place the origin and early
evolution of this extraordinary object in the context of the most
violently star-forming events in the present-day local Universe.

In the past decade, it has become increasingly clear that the most
violently star-forming episodes in the local Universe, such as those
associated with major galaxy mergers and starburst events, produce a
plethora of young massive star clusters (YMCs), often confusingly
referred to as ``super'' star clusters by virtue of their high
luminosities \citep[and references
therein]{whitmore99,degrijs03a,degrijs03b,degrijs03c}.

The issue as to whether at least some of these YMCs might survive for
up to a Hubble time, to eventually evolve into somewhat more
metal-rich counterparts to the ubiquitous GC populations in the local
Universe, has sparked a lively, and ongoing, debate. In essence, the
survival chances of YMCs for any significant length of time depend
crucially on the stellar IMF as well as on environmental factors
\citep[cf.][]{sg01,degrijs05}. If the stellar IMF is either too
shallow (i.e., containing too many high-mass stars with respect to the
low-mass fraction of the population), or characterized by a low-mass
cut-off at stellar masses $m_\star \ga 1$ M$_\odot$, a cluster cannot
survive for more than $\sim 1$ Gyr
\citep[e.g.][]{gnedin97,goodwin97,sg01,mengel02}.

The mere fact that this GC has reached an age of $12.4 \pm 3.2$ Gyr
places useful constraints on its stellar IMF, irrespective of the
cluster's current or initial mass. In particular, it implies that
037-B327's stellar IMF must have had a significant fraction of
low-mass stars, so that IMF descriptions including a low-mass cut-off
above a few M$_\odot$ are ruled out. We cannot, however, distinguish
between the ``standard'' \citet{salpeter55} IMF and more modern IMF
representations that include shallower slopes below $\sim 1$
M$_\odot$.

\begin{table*}
\caption{037-B327 as a ``super'' star cluster.}
\label{ssc.tab}
\begin{center}
\begin{tabular}{ccccc}
\hline
SSP models & IMF$^a$ & Metallicity & $M_V^0(t=10{\rm Myr})$ &
$\mathcal{M}_{\rm GC}$ \\
 & & $(Z)$ & (mag) & ($\times 10^7$ M$_\odot$) \\
\hline
BC03$^b$    & \citet{chabrier03} & 0.0001 & $-16.65^{+0.25}_{-0.22}$ & $2.8^{+0.5}_{-0.7}$ \\
            &                    & 0.0004 & $-16.91 \pm 0.24$        & $2.7^{+0.5}_{-0.8}$ \\
            & \citet{salpeter55} & 0.0001 & $-16.38^{+0.23}_{-0.16}$ & $3.5^{+0.5}_{-0.7}$ \\
            &                    & 0.0004 & $-16.64 \pm 0.20$        & $3.3^{+0.6}_{-0.7}$ \\
{\sc galev} & \citet{salpeter55} & 0.0004 & $-17.14^{+0.24}_{-0.22}$ & $2.7^{+0.5}_{-0.6}$ \\
            & \citet{kroupa01}   & 0.0004 & $-17.07^{+0.25}_{-0.22}$ & $1.5 \pm 0.3$       \\
\hline
\end{tabular}
\end{center}
$^a$ All IMFs were populated from 0.1 to 100 M$_\odot$;
$^b$ Using the 1994 Padova evolutionary tracks.
\end{table*}

Based on its present luminosity, $V = 16.82 \pm 0.05$ mag and very
high extinction, $E(B-V) = 1.36 \pm 0.01$, its intrinsic luminosity,
$V_0 = 12.62 \pm 0.12$ mag [assuming the \citet{car89} Galactic
reddening law; $A_V = 4.20 \pm 0.12$ mag] makes it the intrinsically
most luminous GC in M31 \citep[cf.][]{bk02}.

We will now take this interpretation one step further, by employing a
number of commonly-used SSP models to evolve its luminosity back to a
fiducial age of 10 Myr, so that we can compare its properties with
those expected for YMCs and ``super'' star clusters. In addition, we
can now use the IMF constraint obtained above to bracket the most
likely mass of 037-B327. We list the fiducial absolute $V$-band
magnitudes at 10 Myr, corrected for foreground extinction, as well as
our mass estimates based on a variety of relevant SSP models in Table
\ref{ssc.tab}.

Close inspection of the values for $M_V^0(t = 10\;{\rm Myr})$ shows
that, when it was newly-formed, this cluster truly belonged in the
exceptional class of the ``super'' star clusters. Few, if any, YMCs in
the local Universe exhibit similarly high intrinsic
luminosities. Notable exceptions are NGC 7252-W3 \citep[$M_V^0 \simeq
-16.2$ mag, at a current age of $\sim 3$ Gyr;][]{maraston04}, the
brightest star cluster-like object in NGC 6745 \citep[$M_V^0 \simeq
-13.3$ mag at a current age of $\sim 1$ Gyr;][]{degrijs03c}, as well
as M82-F and a few nuclear star clusters \citep[see for an overview
Table 1 in][]{degrijs05}.

Our mass estimates, $\mathcal{M}_{\rm GC} \ga 2 \times 10^7$
M$_\odot$, place it firmly at the top of the cluster mass function
in the Local Group. \citet{bk02}'s earlier mass estimate of
$\mathcal{M}_{\rm GC} \sim 8.5 \times 10^6$ M$_\odot$ was based on
a photometric mass estimate using a generic mass-to-light (M/L)
ratio, M/L = 2, which was clearly somewhat low for its age. Here
we have shown that, irrespective of the SSP models and stellar IMF
representation assumed, and taking the uncertainties in the
object's metallicity and age into account, the cluster is
significantly more massive than both G1 in M31
\citep[$\mathcal{M}_{\rm G1} = (7-17) \times 10^6$
M$_\odot$;][]{meylan01} and $\omega$ Cen in the Milky Way
\citep[$\mathcal{M}_{\omega{\rm Cen}} = (2.9 - 5.1) \times 10^6$
M$_\odot$;][]{meylan02}, the next most massive clusters (of any
age) in the Local Group\footnote{Even in the unlikely event that
we have overestimated the cluster's age significantly, and that
its true age is $\sim 5$ Gyr, this would reduce the derived
photometric cluster mass by only a factor of $\la 2$, depending on
the metallicity and IMF assumed. Our claim that this cluster is
the most massive star cluster in the Local Group, of any age, is
therefore robust.}.

Because of the key constraints we were able to place on the shape of
the low-mass range of the stellar IMF, and the fact that our
photometric mass estimates are all within each other's uncertainty
ranges [with the exception of the result based on the \citet{kroupa01}
IMF], we predict that dynamical mass estimates will yield very similar
results. Unfortunately, there are no velocity dispersion measurements
for 037-B327 available, however. Using archival {\sl HST}/ACS (Wide
Field Camera) observations in the F606W and F814W filters (programme
GO-10260; PI Harris), we obtained the half-light radii at the
respective wavelengths of these filters. The half-light radii
obtained, corrected (in quadrature) for the intrinsic size of the PSF,
are $(0.52 \pm 0.04)$ arcsec and $(0.53 \pm 0.04)$ arcsec for F606W
and F814W, respectively, corresponding to linear sizes of $(2.52 \pm
0.19)$ pc and $(2.57 \pm 0.19)$ pc, respectively, at the distance of
M31, $m-M = 24.88$ mag. If we combine these half-light radii with the
photometric mass determinations for 037-B327, we predict (using the
virial approximation) that its (one-dimensional) velocity dispersion
will be of order $(72 \pm 13)$ km s$^{-1}$ {\it if} our IMF
assumptions are valid \citep[see, e.g.,][]{maraston04}. Therefore,
spectroscopic observations at a resolution of $R \equiv \lambda /
\Delta \lambda \ga 4200$ (e.g., $\Delta \lambda \ga 2$\AA\ at $\lambda
\sim 8500$\AA, i.e., in the Calcium triplet region) will be able to
confirm our conclusions regarding the low-mass IMF shape of this
GC\footnote{We note that our virial approximation is based on a
generic stellar population of equal-mass stars, and does not take into
account the effects of mass segregation. However, these effects will
affect the predicted velocity dispersion by an amount that is expected
to be within our current uncertainty estimate.}.

We conclude, therefore, that 037-B327 is the current best example of a
``super'' star cluster come of age, and as such an important object
for theorists to take into account when developing models aimed at
describing the evolution of the YMCs seen in large numbers in nearby
starburst and galaxy merger environments.

\section{Summary and conclusions}
\label{summary.sec}

In this paper, we first redetermined the reddening and age of the M31
GC 037-B327 by comparing its independently obtained multicolour
photometry with theoretical stellar population synthesis models. Our
multicolour photometric data are from $UBVRI$, 13 intermediate-band
filters and $JHK{\rm s}$, which constitute an SED covering $\sim
3000-20000$\AA. The reddening towards this cluster, which we determine
at $E(B-V)=1.360 \pm 0.013$ mag, was also estimated by
\citet{bh00,bk02} using the correlations between optical and infrared
colours and metallicity, by defining various ``reddening-free''
parameters, and by using the spectroscopic metallicity to predict the
intrinsic colours. These three different methods yield very consistent
reddening values. The age of $12.4 \pm 3.2$ Gyr for 037-B327 obtained
in this paper shows that 037-B327 is a GC as old as the majority of
the Galactic GCs.

Subsequently, we placed the origin and early evolution of this M31
GC in the context of the YMCs currently being observed in major
starbursts and galaxy mergers in the local Universe. 037-B327 is a
prime example of an unusually bright early counterpart to the
ubiquitous ``super'' star clusters presently observed in most
high-intensity star-forming regions. In order to have survived for
a Hubble time, we conclude that its stellar IMF cannot have been
top-heavy, i.e., characterized by a low-mass cut-off at $m_\star
\ga 1$ M$_\odot$, as sometimes advocated for current ``super''
star clusters. Using this constraint, and a variety of SSP models,
we determine a photometric mass for 037-B327 of $\mathcal{M}_{\rm
GC} = (3.0 \pm 0.5)\times 10^7$ M$_\odot$, somewhat depending on
the SSP models used, the metallicity and age adopted and the IMF
representation. In view of the large number of free parameters,
the uncertainty in our photometric mass estimate is surprisingly
small. This mass, and its relatively small uncertainties, make
this object the most massive star cluster of any age in the Local
Group. Assuming that the photometric mass estimate thus derived is
fairly close to its dynamical mass (based on the assumption of
virial equilibrium), we predict that this GC has a
(one-dimensional) velocity dispersion of order $(72 \pm 13)$ km
s$^{-1}$, which -- if confirmed using spectroscopic observations
at $R \ga 4200$ -- will serve as robust confirmation of our
conclusions regarding the shape of the IMF. As a surviving
``super'' star cluster, this object is of prime importance for
theories aimed at describing massive star cluster evolution.

It has been speculated that some of the most luminous known
globular clusters in the Local Group, including $\omega$ Centauri,
G1-Mayall II and also NGC 2419 \citep[but see][for
counterarguments for this object]{degrijs05}, might be the remnant
nuclei of tidally stripped dwarf galaxies
\citep{zinnecker88,freeman93,bassino94,bergh04}. \citet{ma06}
determined the structural parameters of 037-B327 by fitting the
observed surface brightness distribution to a King profile, and
found that this object falls in the same region of the $M_V$ vs.
log $R_h$ diagram as do $\omega$ Centauri, M54 and NGC 2419 in the
Milky Way and the massive cluster G1 in M31. All four of these
objects are claimed to be the stripped cores of former dwarf
galaxies \citep[see details from][]{bergh04,mackey05}. This
suggests that 037-B327 may also be the stripped core of a former
dwarf companion to M31. However, from ground-based observations of
the brightest objects in NGC 5128, which is the nearest giant
elliptical galaxy, \citet{gomez05} have concluded that clusters
form a continuum in the $M_V$ versus log $R_h$ diagram. So, it is
difficult to distinguish between a globular cluster and a stripped
core of a dwarf. Future work is needed to confirm that 037-B327 is
a globular cluster or whether it might be a stripped core of a
dwarf.

\section*{Acknowledgments}
We acknowledge useful discussions with and suggestions by Nate
Bastian; we are also indebted to the referee for his/her thoughtful
comments and insightful suggestions that improved this paper
greatly. This work has been supported by the Chinese National Key
Basic Research Science Foundation (NKBRSF TG199075402) and by the
Chinese National Natural Science Foundation, No. 10473012 and
10573020.

\end{document}